\documentclass[aps,prd,preprintnumbers,nofootinbib,floatfix,11pt]{revtex4}

\usepackage{amsfonts,amsthm,amsmath,amssymb,mathrsfs,graphicx,epsfig,color,ulem,comment} 
\usepackage{hyperref}
\allowdisplaybreaks[3]
\newcommand{\be}{\begin{equation}}
\newcommand{\ee}{\end{equation}}
\newcommand{\ba}{\begin{eqnarray}}
\newcommand{\ea}{\end{eqnarray}}

\newcommand{\vt}{\vartheta}

\newcommand{\Cr}[1]{{{\cal O} \left( r^{#1} \right)}}
\newcommand{\Crs}[1]{o\left(r_0^{~#1}\right)}
\newcommand{\Crz}[1]{{{\cal O} \left( r_0^{~#1} \right)}}

\newtheorem*{theorem*}{Theorem}
\newtheorem*{proposition*}{Proposition}

\newcommand{\red}[1]{{\textcolor{black}{{#1}}}}

\newtheorem*{lemma*}{Lemma}
\DeclareRobustCommand{\Erase}{\bgroup\markoverwith{\textcolor{red}{\rule[.5ex]{2pt}{0.4pt}}}\ULon}

\begin{document}
\begin{flushright}
  YITP-22-75
  \end{flushright}

\title{Asymptotic behavior of null geodesics near future null infinity. III.\\ Photons towards inward directions}
\author{Masaya Amo$^1$, Keisuke Izumi$^{2,3}$, Yoshimune Tomikawa$^4$,\\ Hirotaka Yoshino$^{5}$ and Tetsuya Shiromizu$^{3,2}$}

\affiliation{$^{1}$Center for Gravitational Physics and Quantum Information, Yukawa Institute for Theoretical Physics, Kyoto University, Kyoto 606-8502, Japan}
\affiliation{$^{2}$Kobayashi-Maskawa Institute, Nagoya University, Nagoya 464-8602, Japan} 
\affiliation{$^{3}$Department of Mathematics, Nagoya University, Nagoya 464-8602, Japan}
\affiliation{$^{4}$Division of Science, School of Science and Engineering, Tokyo Denki University, Saitama 350-0394, Japan}
\affiliation{$^{5}$Department of Physics, Osaka Metropolitan University, Osaka 558-8585, Japan}
\begin{abstract}
  A new sufficient condition for photons emitted near future null infinity to reach future null infinity 
is derived by studying null geodesics
in the Bondi coordinates in asymptotically flat spacetimes.
In our previous works~\cite{Amo:2021gcn,Amo:2021rxr},
such a condition was established for
photons emitted in outward or tangential directions to constant 
radial surfaces. This paper improves our previous result by including photons emitted in inward directions. 
In four dimensions, imposing the same assumptions on the
metric functions as previously,
we prove that photons 
reach future null infinity if their 
initial values of $|dr/du|$ are
smaller than a certain quantity, 
where $r$ and $u$ are the radial and retarded time coordinates, respectively.
This quantity is determined by the asymptotic properties of the metric and is connected to the conjectured maximal luminosity. 
In higher dimensions, photons emitted 
with $dr/du>-(1-1/\sqrt{3})\approx -0.423$
are shown to reach future null infinity 
without the assumptions on the metric functions.
\end{abstract}

\maketitle

%
%
%
\section{Introduction}

Black holes are of great interest both from theoretical and observational points of view. 
The gravitational waves and the image of the black hole were first observed in the 2010s \cite{Abbott:2016blz,Akiyama:2019cqa}, 
which opened new windows to probe our Universe, in particular, the compact objects. 
The theoretical analysis of null geodesics around black holes has provided us plentiful 
knowledge about the observational features, such as the shape of black hole shadows 
(see, e.g. Refs. \cite{Cunha:2017eoe,Cunha:2018acu}).
In particular, in static and spherically symmetric spacetimes, 
circular photon orbits exist and
a set of them is called the photon sphere \cite{Claudel:2000}. 
In less symmetric spacetimes, generalizations of the photon sphere have been 
discussed \cite{Shiromizu:2017ego,Yoshino:2017gqv,Siino:2019vxh,Yoshino:2019dty,Cao:2019vlu,Siino:2021kep}, 
with which photon orbits around compact objects are characterized.

By contrast, the behavior of null geodesics in the region far away from compact objects, where observers  are theoretically considered to be located, has not been well studied in spacetimes without exact symmetries.
In principle, if both the initial position and the tangent vector of the geodesic are given, the orbit is uniquely determined by the geodesic equations in spacetimes with sufficient differentiability. 
Since the asymptotic region of asymptotically flat spacetimes is approximated by the flat spacetime, 
one may think that the behavior of null geodesics would be almost the same as that in the flat spacetime and they definitely reach future null infinity.
However, it is not straightforward to estimate the behavior of geodesics in general asymptotically flat spacetimes without symmetry.


In our previous paper \cite{Amo:2021gcn}, null geodesics corresponding to worldlines of photons initially emitted with non-negative $dr/du$, where $r$ and $u$ are radial coordinate and retarded time, respectively, in asymptotically flat spacetimes were investigated. 
It was shown that in higher dimensions these photons reach future null infinity without imposing assumptions, whereas
in four dimensions several conditions are required for photons to reach future null infinity.
This result suggests that even at sufficiently large $r$, 
the trajectories of photons are not approximated by those in the four-dimensional Minkowski spacetime although the metric asymptotes to that of the Minkowski spacetime. 
Two conditions required in four dimensions correspond to the situation in which the gravitational waves and matter radiations are not so strong near future null infinity.
Although there exist asymptotically flat solutions in which photons can stay on a circular orbit near future null infinity for finite time by quite strong radiation \cite{Amo:2021rxr}, 
photons emitted with non-negative $dr/du$ in the asymptotic region will reach future null infinity in rather realistic spacetimes. 

It is natural to expect that the situation would be similar for photons emitted with negative $dr/du$ but not so large absolute value.  
Motivated by this expectation, in this paper, we analyze orbits of photons emitted with negative $dr/du$ and show the conditions for them to reach future null infinity. 
This analysis is also important in the discussion of the existence of the inner dark horizon and the outer dark horizon~\cite{IDHODH}, which are generalizations of the photon spheres in general asymptotically flat spacetimes
under investigation by the present authors.

The rest of this paper is organized as follows. In Sec.~\ref{review}, we provide a brief overview of the asymptotic behavior 
of the metric near future null infinity in asymptotically flat spacetimes. 
In Sec.~\ref{geo}, we investigate geodesic equations and the null condition at the leading order in the $1/r$ expansion. In Sec.~\ref{proof}, 
we prove that null geodesics corresponding to photons emitted in the direction with negative $dr/du$ but not so large $|dr/du|$ 
reach future null infinity. Section~\ref{Conclusions} is devoted to a summary and discussion.
In the Appendix we consider the outgoing Vaidya spacetimes
without assuming the details of the mass function
and see that the condition on the emitting directions for photons 
to reach future null infinity is relaxed in these spacetimes. We assume the metric to be $C^{2-}$ functions ({\it i.e.}, class $C^{1,1}$) throughout the paper.

%

\section{Brief review of null asymptotics in the Bondi coordinate} 
\label{review}

In this section, we review the asymptotic behavior of the metric near future null infinity in asymptotically flat 
spacetimes based on Refs.~\cite{Bondi,Sachs,Tanabe:2011es} (see also Refs. \cite{Hollands:2003ie,Hollands:2003xp,Ishibashi:2007kb}).

Let $n$ $(\geq 4)$ be the dimension of spacetimes. 
The metric near future null infinity in the Bondi coordinates is written as
\begin{eqnarray}
    \label{Bondi}
ds^2 =g_{\mu\nu}dx^\mu dx^\nu= -Ae^B du^2 -2 e^B du dr + h_{IJ}r^2(dx^I + C^I du)(dx^J + C^J du),
\end{eqnarray}
where the Greek indices denote the spacetime components. Here, $u$ denotes the retarded time, $r$ is 
the areal radius, and $x^I$ stands for the angular coordinates. $A, B, C^I$ and $h_{IJ}$ are functions of $u$, $r$, and $x^I$. 
Future null infinity is described by the limit $r \to \infty$ while $u$ is finite. 
The functions $A$, $B$, $C^I$ and $h_{IJ}$ are expanded
in the power of $1/r$ as\footnote{The Landau symbols of Eq.~\eqref{metric} actually have the higher order by $r^{1/2}$ in even dimensions. We write them as in odd dimensions for unification.} \cite{Tanabe:2011es}
\begin{eqnarray}
    \label{expABC}
A&=&1+ \sum^{k<n/2-2}_{k=0}A^{(k+1)}r^{-(n/2+k-1)}-m(u,x^I)r^{-(n-3)}+\Cr{-(n-5/2)},\\
B&=&B^{(1)}r^{-(n-2)}+\Cr{-(n-3/2)},\\
C^I&=& \sum^{k<n/2-1}_{k=0}C^{(k+1)I}r^{-(n/2+k)}+J^I(u,x^I)r^{-(n-1)}+\Cr{-(n-1/2)},
    \label{fallABCh}\\
    \label{hass}
h_{IJ}&=&\omega_{IJ}+\sum_{k\geq0}h^{(k+1)}_{IJ} r^{-(n/2+k-1)},
\end{eqnarray}
where $\omega_{IJ}$ is the metric for the unit $(n-2)$ sphere, $k\in\mathbb{Z}$ in even dimensions, and $2k\in\mathbb{Z}$ in odd dimensions. 
In particular, in general relativity, the integration of $m(u,x^I)$ over the solid angle provides us the Bondi mass, 
\begin{eqnarray}
M(u):=\frac{n-2}{16\pi}\int_{S^{n-2}}md\Omega \label{M(u)}.
\end{eqnarray}
For $n = 4$, $A^{(1)}$ is set to be zero  because it is absorbed into $m$.
Let us impose the gauge condition   
\begin{align}
    \label{gau}
    \sqrt{\det h_{IJ}}=\omega_{n-2},
  \end{align}
where $\omega_{n-2}$ is the volume element of the unit $(n-2)$-dimensional sphere.
Here, nonzero $h_{IJ}-\omega_{IJ}$ corresponds to the existence
of gravitational waves.
The nonzero components of the metric behave as
\begin{eqnarray}
g_{uu}&=&-Ae^B+h_{IJ}C^IC^J r^2 =-1-A^{(1)}r^{-(n/2-1)}+mr^{-(n-3)} +\Cr{-(n-1)/2}, \nonumber\\
g_{ur} &=& -e^B =-1 -B^{(1)}r^{-(n-2)} + \Cr{-(n-3/2)},\nonumber\\ 
g_{IJ} &=& h_{IJ}r^2 = \omega_{IJ}r^2 + h^{(1)}_{IJ}r^{-(n/2-3)} + \Cr{-(n-5)/2}, \nonumber \\
\quad g_{uI}&=& h_{IJ} C^J r^2 = C^{(1)}_{~~I}r^{-(n/2-2)} +\Cr{-(n-3)/2},\label{metric}
\end{eqnarray}
where $\omega_{IJ}$ denotes the metric of the unit $(n-2)$-dimensional sphere and $C^{(1)}_{~~I}$ is defined as $C^{(1)}_{~~I}:=C^{(1)J}\omega_{IJ}$. 

%
%
%

\section{Geodesic equations and the null conditions}
\label{geo}
In this section, we present the geodesic equations and the null conditions near future null infinity in $n$ $(\geq 4)$ 
dimensions. We define the prime and the dot as the derivative with respect to the affine parameter $\lambda$ and the 
retarded time $u$, respectively. 

By using the metric in Eq.~\eqref{metric}, we can expand the geodesic equations near future null infinity as
\begin{eqnarray}
  {u}'' &=& -\Gamma^{u}_{uu} {u^\prime}^2  -2\Gamma^{u}_{uI} u' \left(x^I\right)^\prime   -\Gamma^{u}_{IJ} \left(x^I\right)^\prime  \left(x^J\right)^\prime \nonumber\\
&=&   -\left[\frac{n-2}{4}A^{(1)}r^{-n/2}+\left(\dot{B}^{(1)}-\frac{n-3}{2}m\right)r^{-(n-2)}+\Cr{-(n+1)/2}\right] {u^\prime}^2  \nonumber\\
&&\hspace{10.5mm}-\left[-\frac{n-4}{2}C^{(1)}_{~~I}r^{-(n/2-1)}+ \Cr{-(n-1)/2}\right] u' \left(x^I\right)^\prime   \nonumber\\
&&\hspace{10.5mm}- \left[\omega_{IJ} r -\frac{n-6}{4}h^{(1)}_{IJ}r^{-(n/2-2)} +\Cr{-(n-3)/2}\right] \left(x^I\right)^\prime  \left(x^J\right)^\prime , \label{eqund}\\
r'' &=&-\Gamma^{r}_{uu}{u^\prime}^2 - 2\Gamma^{r}_{ur}u'r' - \Gamma^{r}_{rr} {r^\prime}^2 - 2\Gamma^{r}_{uI} u' \left(x^I\right)^\prime  - 2\Gamma^{r}_{rI} r' \left(x^I\right)^\prime  - \Gamma^{r}_{IJ} \left(x^I\right)^\prime \left(x^J\right)^\prime  \nonumber\\
 &=& -\left[ \frac{1}{2}\dot{A}^{(1)}r^{-(n/2-1)}-\frac{1}{2}\dot{m}r^{-(n-3)} + \Cr{- (n-1)/2}\right] {u^\prime}^2\nonumber\\
 &&\hspace{8mm}
  -\Big[- \frac{n-2}{2}A^{(1)}r^{-n/2}+(n-3)mr^{-(n-2)}+\Cr{-(n+1)/2}  \Big] u'r'
  \nonumber\\
 &&\hspace{8mm}-\left[-(n-2)B^{(1)}r^{-(n-1)}+\Cr{-(n-1/2)} \right] {r^\prime}^2
\nonumber\\
&&\hspace{8mm}
-\left[\left(\frac{n-4}{2}C^{(1)}_{~~I}+A^{(1)}_{,I}\right)r^{-(n/2-1)}-\left(m_{,I}-C^{(1)J}\dot{h}^{(1)}_{IJ}\right)r^{-(n-3)}\right.  \nonumber \\
&&\hspace{15mm}\left.+\Cr{-(n-1)/2} \right] u' \left(x^I\right)^\prime   \nonumber\\
&&\hspace{8mm}-\left[\frac{n}{2}C^{(1)}_{~~I}r^{-(n/2-1)}+\Cr{-(n-1)/2}\right]r' \left(x^I\right)^\prime \nonumber\\
&&\hspace{8mm}-\left[ -\omega_{IJ} r+\frac{1}{2}\dot{h}^{(1)}_{IJ}r^{-(n/2-3)} +\Cr{-(n-5)/2}\right] \left(x^I\right)^\prime \left(x^J\right)^\prime   , \label{eqr4d}
\\
   \left(x^I\right)'' &=& -\Gamma^{I}_{uu} {u^\prime}^2-2\Gamma^{I}_{ur} u' r'  -2\Gamma^{I}_{uJ} u' \left(x^J\right)^\prime-2\Gamma^{I}_{rJ} r' \left(x^J\right)^\prime   -\Gamma^{I}_{JK} \left(x^J\right)^\prime  \left(x^K\right)^\prime \nonumber\\
&=&
   -\left[\dot{C}^{(1)I}r^{-n/2}+ \Cr{-(n+1)/2}\right] {u^\prime}^2 \nonumber\\
   &&\hspace{10.5mm}-\left[-\frac{n-4}{2}C^{(1)I}r^{-(n/2+1)}+ \Cr{-(n+3)/2}\right]u'r'  \nonumber\\
   &&\hspace{10.5mm}-\left[\dot{h}^{(1)I}_{~~~J}r^{-(n/2-1)}+\Cr{-(n-1)/2}\right] u' \left(x^J\right)^\prime   \nonumber\\
   &&\hspace{10.5mm}-\left[2\delta^I_Jr^{-1} -\frac{n-2}{2}h^{(1)I}_{~~~J}r^{-n/2} +\Cr{-(n+1)/2}\right]r'\left(x^J\right)^\prime \nonumber\\
   &&\hspace{10.5mm}-\Bigg{[}{}^{(\omega)}\Gamma^I_{JK}
   -\left\{C^{(1)I}\omega_{JK}-\frac{1}{2}\left(D_Jh^{(1)I}_{~~~K}+D_Kh^{(1)I}_{~~~J}-D^Ih^{(1)}_{~JK}\right)\right\}r^{-(n/2-1)}\nonumber \\
   &&\hspace{15mm}+\Cr{-(n-1)/2}\Bigg] \left(x^J\right)^\prime  \left(x^K\right)^\prime,\label{eqI4d}
\end{eqnarray}
\red{where the subscript 
 ``$_{,I}$'' is the derivative with respect to $x^I$, ${}^{(\omega)}\Gamma^I_{JK}$ is defined as ${}^{(\omega)}\Gamma^I_{JK}:=\frac{1}{2}\omega^{IL}\left(\omega_{JL,K}+\omega_{KL,J}-\omega_{JK,L}\right)$,} and $D_I$ is the covariant derivative with respect to $\omega_{IJ}$.

The condition for the geodesic to be null is 
\ba
    {u^\prime}^2&=&-2\left[1-A^{(1)}r^{-(n/2-1)}+mr^{-(n-3)}+\Cr{-(n-1)/2}\right]u'r'\nonumber\\
    &&\hspace{4mm}+\left[\omega_{IJ}r^2+\left(h^{(1)}_{IJ}-A^{(1)}\omega_{IJ}\right)r^{-(n/2-3)}+m\omega_{IJ}r^{-(n-5)}+\Cr{-(n-5)/2}\right]\left(x^I\right)^\prime \left(x^J\right)^\prime \nonumber\\
    &&\hspace{4mm}+\left[2C^{(1)}_{~~I}r^{-(n/2-2)}+\Cr{-(n-3)/2}\right]\left(x^I\right)^\prime u' \nonumber \\
    &=&-2\Big[1+\Cr{-(n/2-1)}\Big]u'r'+\left[r^2+\Cr{-(n/2-3)}\right]\left|\left(x^I\right)^\prime\right|^2 +\Cr{-(n/2-1)}u'^2,\label{eqnull}
\ea
where $\left|\left(x^I\right)^\prime \right|$ was introduced as 
    \begin{eqnarray}
    \left|\left(x^I\right)^\prime \right| := \sqrt{\omega_{IJ} \left(x^I\right)^\prime  \left(x^J\right)^\prime }
    \end{eqnarray}
and in the second equality we used 
\ba
\left|\left(x^I\right)^\prime\right|\left| u'\right|=\Cr{1}\left|\left(x^I\right)^\prime\right|^2+\Cr{-1}\left|u^\prime\right|^{2}\red{.}
\ea
\red{This} can be confirmed by using the arithmetic-geometric mean inequality
\red{
\ba
 \left|\left(x^I\right)^\prime \right| \left|u'\right|\leq  \frac12 r\left|\left(x^I\right)^\prime \right|^2+\frac12 r^{-1}\left|u'\right|^2
\ea 
and the definition of the Landau symbol.}
Equation \eqref{eqnull} gives us
\ba
\left|\left(x^I\right)^\prime\right|^2 &=&2\Big[r^{-2}+\Cr{-(n/2+1)}\Big]u'r'+\left[r^{-2}+\Cr{-(n/2+1)}\right]{u^\prime}^2\label{nullIndim}.
\ea
In four dimensions, this equation becomes
\ba
\left|\left(x^I\right)^\prime\right|^2 &=&2\Big[r^{-2}+\Cr{-3}\Big]u'r'+\Big[r^{-2}+\Cr{-3}\Big]{u^\prime}^2\label{nullI}.
\ea
Equation \eqref{eqnull} has two solutions for $u'$. We choose the future directed one,
\ba
\hspace{0mm}0\leq u'&=&\left[1+\Cr{-(n/2-1)}\right]\left[-\left\{1+\Cr{-(n-2)}\right\}r'+\Cr{-(n/2-2)}\left| \left(x^I\right)' \right|\right.\nonumber\\
    &&\left.\hspace{1mm}+\sqrt{\left[\left\{1+\Cr{-(n-2)}\right\}r'+\Cr{-(n/2-2)}\left| \left(x^I\right)' \right|\right]^2+\left[r^2+\Cr{-(n/2-3)}\right]\left| \left(x^I\right)' \right|^2}\right]\nonumber\\
    &\leq&-\left[1+\Cr{-(n/2-1)}\right]r'+\Cr{-(n/2-2)}\left| \left(x^I\right)' \right|+\left[1+\Cr{-(n-2)}\right]r'\nonumber\\
    &&\hspace{12mm}+\Cr{-(n/2-2)}\left| \left(x^I\right)' \right|+\left[r+\Cr{-(n/2-2)}\right]\left|\left(x^I\right)'\right|
    \nonumber\\
    &=&\left[r+\Cr{-(n/2-2)}\right]\left|\left(x^I\right)'\right|+\Cr{-(n/2-1)}r',\label{ApCnullcon1}
    \ea
    where we used the formula $\sqrt{a+b}\leq\sqrt{a}+\sqrt{b}$ which is valid for $a\geq0$ and $b\geq0$.
By substituting Eq.~\eqref{ApCnullcon1} into  Eq.~\eqref{eqr4d}, we have
\ba
r'' 
 &=& -\left[ \frac{1}{2}\dot{A}^{(1)}r^{-(n/2-1)}-\frac{1}{2}\dot{m}r^{-(n-3)} + \Cr{- (n-1)/2}\right] {u^\prime}^2+\Cr{-n/2}u'r'\nonumber\\
 &&\hspace{8mm}
  +\Cr{-(n/2-1)} u' \left| \left(x^I\right)' \right| +\Cr{-(n-1)}{r^\prime}^2+\Cr{-(n/2-1)}r' \left| \left(x^I\right)' \right| \nonumber\\
&&\hspace{8mm}-\left[ -\omega_{IJ} r+\frac{1}{2}\dot{h}^{(1)}_{IJ}r^{-(n/2-3)} +\Cr{-(n-5)/2}\right] \left(x^I\right)^\prime \left(x^J\right)^\prime  \nonumber\\
&=&\left(\omega_{IJ}-\frac{1}{2}\dot{h}^{(1)}_{IJ}r^{-(n/2-2)}-\frac{1}{2}\dot{A}^{(1)}\omega_{IJ}r^{-(n/2-2)}+\frac{1}{2}\dot{m}\omega_{IJ}r^{-(n-4)}\right)r\left(x^I\right)'\left(x^J\right)'\nonumber\\
&&\hspace{4mm}+\Cr{-(n-5)/2}\left|\left(x^I\right)'\right|^2+\Cr{-(n-1)}r'^2+\Cr{-(n/2-1)} r' \left| \left(x^I\right)' \right|.
\label{ApCr''high}
\ea
All of the terms in the bracket of the first term after the last equality of Eq.~\eqref{ApCr''high} are 
the same order in four dimensions,
while only $\omega_{IJ}$ is dominant in dimensions higher than four. 
It brings an essential difference to the asymptotic behavior of null geodesics between four and higher dimensions. 
In particular, as seen in our previous paper~\cite{Amo:2021gcn}, in four dimensions, such contributions might hinder 
asymptotic  null geodesics from getting to future null infinity. In four dimensions, 
we define
\begin{equation}
  \Omega_{IJ} :=  \omega_{IJ}  - \frac12 \dot{h}^{(1)}_{IJ} + \frac12 \dot{m} \omega_{IJ},
  \label{Def:Omega_IJ}
\end{equation}
and then 
Eq.~\eqref{ApCr''high} becomes
\ba
r'' &=&\Omega_{IJ}r\left(x^I\right)'\left(x^J\right)'+\Cr{0}\left|\left(x^I\right)'\right|^2+\Cr{-3}r'^2+\Cr{-1} r' \left(x^I\right)^\prime\nonumber\\
&=&\Omega_{IJ}r\left(x^I\right)'\left(x^J\right)'+\Cr{0}\left|\left(x^I\right)'\right|^2+\Cr{-2}r'^2,\label{ApCr''}
\ea
where we used the arithmetic-geometric mean inequality
\red{
\ba
\left|r'\right| \left|\left(x^I\right)^\prime \right|\leq  \frac12 r^{-1}\left|r'\right|^2 + \frac12 r\left|\left(x^I\right)^\prime \right|^2
\ea}in the last equality. 
As seen in our previous paper~\cite{Amo:2021rxr}, $\Omega_{IJ}$ is related to the extrinsic curvature of $r$-constant 
hypersurfaces and its vanishing means the existence of approximate photon surfaces near future null infinity. 

%
\section{Null geodesics to reach future null infinity}
\label{proof}

The authors of this paper have shown that any photon emitted from near future null infinity with 
$\dot{r}\geq0$ reaches future null infinity, if $\Omega_{IJ}$ is positive definite and $\dot{m}\leq0$ holds~\cite{Amo:2021gcn} 
(see Ref. \cite{Cao:2021mwx} for the extension to the Brans-Dicke theory). Our interest in the present paper is that
the lower bound of the initial value of $\dot{r}$
would not have to be strictly zero, and we will show in this section that it can be negative. 
We mainly focus on four-dimensional cases, because the generalization to higher-dimensional cases is easy. 

Let us define a function $\Omega(u, {x^I}):=\Omega_{IJ} (u, {x^K})  \phi^I \phi^J$, 
where $\phi^I$ is a vector satisfying $|\phi^I|:= \sqrt{\omega_{IJ} \phi^I\phi^J }=1$. 
In addition, we introduce $\Omega_{i}$ as the infimum of $\Omega$. 
When we assume $\Omega_{IJ}$ to be positive definite as imposed in the Proposition, 
we see that $\Omega_{i}\geq0$ holds. 

As a preparation, let us show that $u'(\lambda)\neq0$ holds for the situation $r^\prime(\lambda)\le 0$. Suppose, for the sake of contradiction, that $u'\left(\lambda_0\right)=0$ and $r'\left(\lambda_0\right)\le0$ hold true at some affine parameter $\lambda_0$. Then, $u'\left(\lambda_0\right)=0$ and Eq. \eqref{nullI} provide $\left|\left(x^I\right)^\prime\left(\lambda_0\right)\right|=0$. By substituting $u'\left(\lambda_0\right)=0$ and $\left|\left(x^I\right)^\prime\left(\lambda_0\right)\right|=0$ to the first equality of Eq. \eqref{ApCnullcon1}, we obtain
\ba
\left[-2+\Cr{-1}\right]r'\left(\lambda_0\right)=0,
\ea 
where we used the negativity of $r'\left(\lambda_0\right)$. This provides us $r'\left(\lambda_0\right)=0$. Thus, all of the components of $\left(x^\mu\right)'$ vanish. Then, $\left(x^\mu\right)'$ is not appropriate as a tangent vector of the null geodesic. Therefore, $u'(\lambda)\neq0$ holds for the situation $r'(\lambda)\le0$,
which is the case we will focus on. Then, we can take the retarded time $u$ as the parameter for the null geodesic. This makes the analysis simpler than the previous study \cite{Amo:2021gcn}.


The main statement of this paper is explicitly written as follows: 
\begin{proposition*}
  \label{lemma3}
Consider a four-dimensional asymptotically flat spacetime in which the metric near future null infinity can be written as Eq.~\eqref{metric} with the Bondi coordinates by $C^{2-}$ functions. 
Suppose that $\Omega_{IJ} :=  \omega_{IJ}  - \frac12 \dot{h}^{(1)}_{IJ} + \frac12 \dot{m} \omega_{IJ}$ is positive definite and $\dot{m}\leq0$ holds everywhere near future null infinity. 
Take a point $p$ with sufficiently large radial coordinate value $r=r_0$.
Then any photon emitted from $p$ reaches future null infinity if 
\ba
0<\frac{1}{\dot{r}|_p-\dot{r}_{\rm crit}}=\Crs{1}\label{con1}
\ea
is satisfied, where $\dot{r}_{\rm crit}$ is defined by 
\ba
\dot{r}_{\rm crit}:=\frac{-\left(2\Omega_i+1\right)+\sqrt{4\Omega_i^2-2\Omega_i+1}}{3}\leq0. \label{def_r_crit}
\ea
\end{proposition*}
\noindent 

We set $u$ to be zero at $p$, which gives us $r(0)=r_0$. 
Then, Eq.~\eqref{con1} means that $\dot{r}(0)>\dot{r}_{\rm crit}$ and that $\dot{r}(0)-\dot{r}_{\rm crit}$ is not too small.
Since we can easily check that 
the right-hand side of Eq. (\ref{def_r_crit}) is a downwardly convex and
monotonically decreasing function of $\Omega_i$ 
for $\Omega_i \ge 0$, 
\red{we see that the graph of $\dot{r}_{\rm crit}=\dot{r}_{\rm crit}(\Omega_i)$, as a function of $\Omega_i$, is softly bounded from below by its tangential line of $\dot{r}_{\rm crit}(\Omega_i)$ at $\Omega_i=0$ and the asymptotic line of $\dot{r}_{\rm crit}(\Omega_i)$ for $\Omega_i \to \infty$}, which gives
\ba
\dot{r}_{\rm crit}& > &-\frac{1}{2},\label{tan1}\\
\dot{r}_{\rm crit}& \geq &-\Omega_i\label{tan2}
\ea
for any $\Omega_i\geq0$.
Therefore, Eq.~\eqref{con1} gives us
\ba
\dot{r}(0)&>&-\red{\frac{1}{2}},\label{-1/2}\\
\dot{r}(0)&>&-\Omega_i.\label{-O}
\ea
In exactly flat spacetime, $-1/2=\dot r (= dr/du)$ with a null condition gives $dx^I/du=0$, which means 
the null geodesics \red{are towards the exact inward direction} satisfying $t=-r+$ const.  
Thus, Eq.~\eqref{-1/2} implies that the initial direction is not \red{towards the exact inward direction}. 
In addition, as we will see later, Eq.~\eqref{-O} avails to show that $r$ is kept large.%
\footnote{In specific spacetimes, the condition to reach future null infinity can be
relaxed further. See the Appendix for the analysis of the outgoing Vaidya spacetime for such an example.}

For $\dot{r}\ge0$, the proof has been already done in our first paper of this series~\cite{Amo:2021gcn}. 
In addition, in the case with $\Omega_i=0$, the condition $\dot{r}>\dot{r}_{\rm crit}$ gives $\dot{r}>0$.
Therefore, only the cases with negative $\dot{r}(0)$ and $\Omega_i>0$ (that is, $\dot{r}_{\rm crit}< \dot{r}<0$) are a remaining issue.
Let us prove that any null geodesic reaches future null infinity in this case.

Suppose there exists the infimum $u_{\rm inf}$ of $u>0$ without satisfying at least 
one of\footnote{The right-hand side of Eq.~\eqref{rLarge} can be replaced by a multiplication of the 
original one by any positive number strictly smaller than ${3}/{2}$ to prove the Proposition.} 
\ba
\dot{r}_{\rm crit}&<&\dot{r}(u)<0,\\
r(u)&\geq& r_0 \left[\dot{r}(0)-\dot{r}_{\rm crit}\right] \left[\dot{r}(0)-\dot{r}_{\rm crit}+\frac23 \sqrt{4\Omega_i^2-2\Omega_i+1}\right] \left[\Omega_{i}+\dot{r}(0)\right]^{-1}\left[2\dot{r}(0)+1\right]^{-1}.\label{rLarge}
\ea
From Eqs.~\eqref{con1}, \eqref{-1/2}, and \eqref{-O}, we find that the right-hand side of 
Eq.~\eqref{rLarge} is $\Crz{1}$. This means that, for sufficiently large $r_0$, 
$r(u)$ is kept sufficiently large. We now are ready to show the strategy of the proof. 
We have introduced the infimum $u_{\rm inf}(<\infty)$ of a range of $u$ where at least one of $\dot{r}_{\rm crit}<\dot{r}(u)$, $\dot{r}(u)<0$ and Eq.~\eqref{rLarge} is violated.
Then, we will examine which of them is violated first. 
Actually, we can show the existence of $u_{\rm inf}(<\infty)$, and prove that $\dot{r}_{\rm crit}<\dot{r}(u)$ and Eq.~\eqref{rLarge} still hold true sufficiently near $u_{\rm inf}$. 
This means that the violation of $\dot{r}(u)<0$ occurs first, that is, $\dot{r}(u)\ge 0$ is achieved at large $r(u)$.
Then we apply 
our previous result~\cite{Amo:2021gcn} to the current issue.

Let us show first that $\dot{r}_{\rm crit}<\dot{r}(u)$ holds for $(0<)u<u_{\rm inf}$, and that $\dot{r}_{\rm crit}<\dot{r}(u)$ still holds even for $u$ slightly larger than $u_{\rm inf}$.
In four dimensions, Eq.~\eqref{eqund} becomes
\ba
{u}'' 
&=&
   \Cr{-2} {u^\prime}^2  +\Cr{-2} u' \left|\left(x^I\right)'\right|  -\left[ r +\Cr{0}\right] \left|\left(x^I\right)'\right|^2 \nonumber\\
   &=&\Cr{-2} {u^\prime}^2   -\left[  r +\Cr{0}\right] \left|\left(x^I\right)'\right|^2 ,\label{equ4d}
\ea
\red{where we used the arithmetic-geometric mean inequality
in the second equality.}
By substituting Eq.~\eqref{nullI} into Eq.~\eqref{equ4d}, we have
\ba
{u}'' 
   &=& -2 \Big[r^{-1}+\Cr{-2}\Big]u'r'- \Big[r^{-1}+\Cr{-2}\Big]{u^\prime}^2.\label{equnull}
\ea
With this equation, $r''$ is written as
\ba
r''&=&\left(u'\dot{r}\right)'\nonumber\\
&=&u''\dot{r}+u'^2\ddot{r}\nonumber\\
&=&\left[-2 \Big\{r^{-1}+\Cr{-2}\Big\}u'r'- \Big\{r^{-1}+\Cr{-2}\Big\}{u^\prime}^2\right]\dot{r}+u'^2\ddot{r}\nonumber\\
&=&\left[-2 \Big\{r^{-1}+\Cr{-2}\Big\}\dot{r}^2- \Big\{r^{-1}+\Cr{-2}\Big\}\dot{r}+\ddot{r}\right]u'^2.\label{r''dot}
\ea
The comparison between Eqs.~\eqref{ApCr''} and \eqref{r''dot} gives
\ba
&&\left[-2 \Big\{r^{-1}+\Cr{-2}\Big\}\dot{r}^2- \Big\{r^{-1}+\Cr{-2}\Big\}\dot{r}+\ddot{r}\right]u'^2\nonumber\\
&&\hspace{30mm}=
\Omega_{IJ}r\left(x^I\right)'\left(x^J\right)'+\Cr{0}\left|\left(x^I\right)'\right|^2+\Cr{-2}r'^2\nonumber\\
&&\hspace{30mm}=\Big[ \Omega_{IJ}r\dot{x}^I\dot{x}^J +\Cr{0}\left|\dot{x}^I\right|^2 +\Cr{-2}\dot{r}^2 \Big] u'^2.\label{rgeoequ}
\ea
By dividing Eq.~\eqref{rgeoequ} by $u'^2$, which is nonzero, we have
\ba
\ddot{r}&=&\Omega_{IJ}r\dot{x}^I\dot{x}^J+\Cr{0}\left|\dot{x}^I\right|^2+2 \Big[r^{-1}+\Cr{-2}\Big]\dot{r}^2+\Big[r^{-1}+\Cr{-2}\Big]\dot{r}\nonumber\\
&\geq&\Omega_{i}\Big[r+\Cr{0}\Big]\left|\dot{x}^I\right|^2+2 \Big[r^{-1}+\Cr{-2}\Big]\dot{r}^2+\Big[r^{-1}+\Cr{-2}\Big]\dot{r}.\label{rdd}
\ea
Similarly, Eq.~\eqref{nullI} gives us
\ba
\left|\dot{x}^I\right|^2 &=&2\Big[r^{-2}+\Cr{-3}\Big]\dot{r}+\Big[r^{-2}+\Cr{-3}\Big]\label{nullIu}.
\ea
\red{For $0\leq u<u_{\rm inf}$, by using $\dot{r}_{\rm crit}<\dot{r}(u)$, Eqs.~\eqref{tan1} and \eqref{tan2}, we have
\ba
\dot{r}(u)&>&-\frac{1}{2},\label{u_-1/2}\\
\dot{r}(u)&>&-\Omega_i.\label{u_-O}
\ea}
By substituting Eq.~\eqref{nullIu} into Eq.~\eqref{rdd}, for $0\leq u<u_{\rm inf}$, we have 
\ba
\ddot{r}
&\geq&\Omega_{i}\left[2\Big\{r^{-1}+\Cr{-2}\Big\}\dot{r}+\Big\{r^{-1}+\Cr{-2}\Big\}\right]+2 \Big[r^{-1}+\Cr{-2}\Big]\dot{r}^2+\Big[r^{-1}+\Cr{-2}\Big]\dot{r}\hspace{4mm}\nonumber\\
&=&\left(\Omega_{i}+\dot{r}\right)\left[2\Big\{r^{-1}+\Cr{-2}\Big\}\dot{r}+\Big\{r^{-1}+\Cr{-2}\Big\}\right]\hspace{4mm}\nonumber\\
&>&0,\label{rdd2}
\ea
where, in the third line, we used \red{Eqs.~\eqref{u_-1/2}, \eqref{u_-O}} and the fact that $r$ is kept sufficiently large for $0\leq u<u_{\rm inf}$ as the consequence of Eq.~\eqref{rLarge} for sufficiently large $r_0$. 
The positivity of $\ddot r$ gives $\dot{r}_{\rm crit}<\dot{r}(0)<\dot{r}(u)$ for $0<u<u_{\rm inf}$.
Recalling that $u'\neq0$ if $r'\le0$, the continuity of $u'$ provides that it holds also for $u$ slightly larger than $u_{\rm inf}$. This gives us the continuity of $\dot{r} (u)$ near $u=u_{\rm inf}$, which yields $\dot{r}_{\rm crit}<\dot{r}(u)$ even for $u$ slightly larger than $u_{\rm inf}$.
Therefore, the condition $\dot{r}_{\rm crit}<\dot{r}(u)$ is irrelevant to
the existence of $u_{\rm inf}$. 
In other words, if $u_{\rm inf}(<\infty)$ exists, either  $\dot{r}<0$ or Eq.~\eqref{rLarge} must be violated at $u=u_{\rm inf}$.

Next, let us show that Eq.~\eqref{rLarge} is still valid near $u=u_{\rm inf}$.
For $0\leq u<u_{\rm inf}$,
the double integral of Eq.~\eqref{rdd2} gives
\ba
r(u)&\geq&r_0+\int^{u}_0 du_1\left[\dot{r}(0)+\int^{u_1}_0 du_2\left[\left(\Omega_{i}+\dot{r}\right)\left\{2\Big(r^{-1}+\Cr{-2}\Big)\dot{r}+\Big(r^{-1}+\Cr{-2}\Big)\right\}\right]\right]\nonumber\\
&\geq &r_0+\int^{u}_0 du_1\left[\dot{r}(0)+u_1\left\{\left(\Omega_{i}+\dot{r}(0)\right)\left(2\dot{r}(0)+1\right)\Big(r_0^{-1}+\Crz{-2}\Big)\right\}\right]\nonumber\\
&=&r_0+u\dot{r}(0)+\frac{1}{2}u^2\left(\Omega_{i}+\dot{r}(0)\right) \left( 2\dot{r}(0)+1\right) \left( r_0^{-1}+\Crz{-2}\right) \nonumber\\
&\geq&r_0-\frac{1}{2}r_0\dot{r}(0)^2\left( \Omega_{i}+\dot{r}(0)\right)^{-1} \left( 2\dot{r}(0)+1\right)^{-1} \Big( 1+\Crz{-1}\Big), \label{r(u)1}
\ea
where we used Eqs.~\eqref{-1/2}, \eqref{-O} and $r(u)\leq r_0$ for $0\leq u<u_{\rm inf}$ in the second line. 
The last line represents the minimum value
of the third line as a function of $u$. 
By  using the
definition of $r_{\rm crit}$ given in Eq.~\eqref{def_r_crit}, we can rewrite the last line of Eq.~\eqref{r(u)1} in
\ba
r(u)&\geq &
\frac{3}{2} r_0 \left[\dot{r}(0)-\dot{r}_{\rm crit}\right] \left[\dot{r}(0)-\dot{r}_{\rm crit}+\frac23 \sqrt{4\Omega_i^2-2\Omega_i+1}\right] \left[\Omega_{i}+\dot{r}(0)\right]^{-1}\left[2\dot{r}(0)+1\right]^{-1}\label{r(u)2}
\ea
for $0\leq u<u_{\rm inf}$. 
From Eqs.~\eqref{con1}, \eqref{-1/2}, \eqref{-O} and the assumption that $r_0$ is large enough,
we see that Eq.~\eqref{rLarge} holds near $u=u_{\rm inf}$ by the continuity and monotonicity of $r(u)$.

Our remaining task is to show the finiteness of $u_{\rm inf}$ to guarantee that $\dot{r}(u) <0$ is actually violated.
For the sake of contradiction, suppose that\footnote{The coefficient 3 in Eq. \eqref{utilde} can be replaced by any positive number strictly larger than 2.}
\ba
u_{\rm inf} > 3r_0\left(-\dot{r}(0)\right) \left( \Omega_{i}+\dot{r}(0)\right)^{-1} \left( 2\dot{r}(0)+1\right)^{-1}=: \tilde u\label{utilde}
\ea
holds true.
For $u=\tilde{u} <u_{\rm inf}$, on the one hand, the second to the last line of Eq.~\eqref{r(u)1} gives $r(\tilde u) >r(0)$. 
On the other hand, however, since $\dot r<0$ holds for $0\leq u <u_{\rm inf}$, $r(u) \leq r(0)$ is satisfied. 
These results contradict each other. 
Thus, $u_{\rm inf} (\le \tilde u <\infty)$ should exist. 
Therefore, $\dot{r}(u)<0$ is violated at some finite $u>0$, and there Eq.~\eqref{rLarge} shows that $r(u)$ is large enough. 

It has been shown in \cite{Amo:2021gcn} that any photon emitted with $\dot{r}(u)\geq0$ from sufficiently large $r(u)$ 
region will reach future null infinity when $\Omega_{IJ}$ is positive definite and $\dot{m}\leq0$ holds. 
Therefore, photons which we consider here reach future null infinity. 
This completes the proof of the Proposition.

It is straightforward to extend the Proposition to the case of higher dimensions. 
In higher-dimensional cases, the leading order contribution of the first term in the right-hand side of Eq.~\eqref{ApCr''high} is only $\omega_{IJ}$, that is, no higher order terms of the metric functions show up.
Therefore, the behavior of null geodesics can be approximated by that in the flat spacetime very well. 
The proof for higher-dimensional cases is obtained by replacing $\Omega_{IJ}$ in this section by $\omega_{IJ}$, which is positive definite.
Thus, photons emitted with $\dot{r}(\lambda)\geq0$ from a sufficiently large $r$ region will reach future null infinity without these conditions as shown in \cite{Amo:2021gcn}. 
Then, the condition to reach future null infinity in higher dimensions is given by Eq.~\eqref{con1} in which 
$\Omega_{i}$ is replaced by 1, {\it i.e.} $\dot{r}_{\rm crit}=-(1-1/\sqrt{3}) \approx -0.423$. 
This result implies that the initial direction should not be so close to the exact inward direction for which $\dot{r}(0)\simeq-0.5$.

In the Appendix, we study the behavior of null geodesics in the four-dimensional outgoing Vaidya spacetimes
without assuming the details of the mass function. 
In such specific spacetimes, the conditions required for the null geodesics to reach future null infinity are relaxed.

%

\section{Summary and Discussion}
\label{Conclusions}

In this paper, we have investigated null geodesics that correspond to photons emitted in inward directions in 
asymptotically flat spacetimes and have discussed whether they reach future null infinity. 
In four dimensions, photons emitted with $\dot{r}\gtrsim \dot{r}_{\rm crit}$, where $\dot{r}_{\rm crit}$ is determined in terms of $\Omega_{IJ}$ as Eq.~\eqref{def_r_crit}, reach future null infinity when $\Omega_{IJ} $ is positive definite and $\dot{m}\leq0$.  
By contrast, in higher dimensions, photons emitted with $\dot{r}\gtrsim\dot{r}_{\rm crit} \approx -0.423$ 
reach future null infinity in general. 
Note that for the initial condition with negative $\dot{r}$ whose absolute value is sufficiently large, 
which is out of the applicability of our Proposition, photons have the possibility to be absorbed into the black hole(s) at the center 
because photons are emitted almost in the central direction
for such initial conditions.

The required conditions for the metric 
in the four-dimensional case are the same as the
ones that were imposed to prove that photons with the initial
condition $\dot{r}(0)\geq0$ reach future null infinity
in our previous paper~\cite{Amo:2021gcn}.
The result of this paper confirms the importance of $\Omega_{IJ}$ and $\dot{m}$ for the asymptotic behavior of null geodesics near future null infinity.
It may be also interesting to look for examples of spacetimes violating the
positivity condition of $\Omega_{IJ}$ in which photons emitted with $\dot{r}\gtrsim\dot{r}_{\rm crit}$ do not reach future null infinity.
Unfortunately, as shown in the Appendix, such an example cannot be realized by 
the simple spherically symmetric Vaidya spacetime. Spacetimes
with strong gravitational waves might have the possibility to
realize such a situation.

Readers might suspect that 
it would be possible to prove the Proposition simply by adopting
the coordinate transformation $\tilde{x}^\mu=x^\mu+\xi^\mu$ 
from the current coordinates
that realizes the initial condition $d\tilde{r}/d\tilde{u}\geq0$
for a photon and by applying our previous result of Ref.~\cite{Amo:2021gcn}. 
However, this method seems to be unrealistic because we need to shift
the position of the coordinate origin by the amount of $\Crz{1}$ in general, and after that the $r$--constant surface
that includes the emission point might not be sufficiently far away compared to the distances to the gravitational sources. This means that the asymptotic form of the Bondi metric may not be applied.

Based on the features that we have found for null geodesics near future null infinity, we can define new geometrical concepts as 
the generalizations of the photon sphere~\cite{IDHODH}, which
is our ongoing work. 
There, in order to show the existence of 
the newly defined generalized photon sphere,
the condition for $\Omega_{IJ}$ and $\dot{m}$
imposed in the present paper 
will play an important role.

It should be noted here that the positive definiteness of
$\Omega_{IJ}$ is closely related to the maximum possible luminosity
of exploding objects.  
In spherically symmetric cases, the condition $\Omega_{IJ}=0$ is equivalent to  $\dot{m}(u)=-2$, and 
substitution of this formula into Eq.~\eqref{M(u)} leads to 
$\dot{M}(u)c^2= -c^5/G$.
This quantity is called the Planck luminosity~\cite{dyson},
and it is conjectured to be the maximum possible luminosity in the Universe~\cite{thorne} (see also discussion of Refs.~\cite{Cao:2021mwx,Amo:2021rxr} for more details). 
The effects of $\Omega_{IJ}$ and/or $\dot{m}$ would give interesting influence on observations performed in asymptotic regions and theoretical studies of asymptotic properties of spacetimes. 

%
%

\acknowledgments

M. A. is grateful to Professor S. Mukohyama and Professor T. Tanaka for continuous encouragements and useful suggestions. 
M. A. is supported by Grant-in-Aid for JSPS Fellows No. 22J20147.
K. I.  and T. S. are supported by Grant-Aid for Scientific Research from Ministry of Education, Science, Sports 
and Culture of Japan (No. JP17H01091 and No. JP21H05182). K. I., H. Y. and T. S. are also supported by JSPS(No. JP21H05189). 
K.~I. is also supported by 
JSPS Grants-in-Aid for Scientific Research (B) (No. JP20H01902)
and 
JSPS Bilateral Joint Research Projects (JSPS-DST collaboration) (No. JPJSBP120227705).
H. Y. is in part supported by JSPS KAKENHI Grant No.  JP22H01220,
and is partly supported by Osaka Central Advanced Mathematical Institute 
(MEXT Joint Usage/Research Center on Mathematics and Theoretical Physics No. JPMXP0619217849).
T. S. is also supported by JSPS Grants-in-Aid for Scientific Research (C) (No. JP21K03551). 

\appendix
%

\section{Spherically symmetric case}
\label{app.a}

In this Appendix, we investigate the behavior of a photon in spherically symmetric cases.
We adopt the four-dimensional outgoing Vaidya metric, which describes a spherically symmetric, dynamical spacetime with outgoing flow of null matter, as a specific example, and estimate the deflection angle of a photon which passes through the null matter domain. 
We will find that in this spacetime, 
a photon emitted near future null infinity
will reach future null infinity
in a fairly generic initial condition.

The metric of the outgoing Vaidya spacetime is
\ba
  ds^2 = -f(u,r)du^2-2dudr+r^2\omega_{IJ}dx^Idx^J,\label{Vaidya}
\ea
with
\begin{equation}
  f(u,r)=1-\frac{2M(u)}{r},
\end{equation}
where $M(u)$ is the Bondi mass. We use the standard
spherical-polar coordinates $(\theta,\phi)$ on the unit sphere.
As for the mass function $M(u)$,
we consider
the situation where
\begin{equation}
  M(u)\ = \ \left\{
  \begin{array}{ll}
    M_0 & (u\le 0),\\
    M_0+\Delta M & (\Delta u \le u)
  \end{array}
  \right.
\end{equation}
and $M(u)$ is monotonically decreasing in the range $0\le u\le \Delta u$
(therefore, $\Delta M <0$).   
Physically, $\Delta u$ represents the duration of the null-matter flow,
and $|\Delta M|$ is the total energy carried by it.  
In this Appendix, we will extract the behavior of a photon
that does not depend on the detailed form of $M(u)$
in the range $0\le u\le \Delta u$.
Also, we do not assume the positive definiteness of $\Omega_{IJ}$.
Throughout this Appendix, $\Delta X$ means the change
in the quantity $X(u)$ from $u=0$ to $u=\Delta u$ [{\it i.e.},
$\Delta X=X(\Delta u)-X(0)$]. 
As written in Sec.~\ref{geo}, the prime and the dot denote the derivative with respect to the affine parameter $\lambda$ and the retarded time $u$, respectively.

Without loss of generality,
we restrict our attention to the case that a photon is emitted at
$u=0$ and $r=r_0$ with
${\phi}^\prime>0$ on the $\theta=\pi/2$ plane.
In this study, the initial radial position of a photon
$r_0$ is regarded as a parameter
to specify the order of physical quantities.
The quantities $\Delta u$ and $M_0$ (as well as $\Delta M$)
are treated as constants that do not depend on $r_0$, and
hence, these quantities are regarded as 
$\Crz{0}$.
The situations considered here are
the cases of $\mathrm{max}[M_0, \Delta u] \ll r_0$,
and include both of the cases $M_0\ll \Delta u$ and $\Delta u\ll M_0$.

We now define the deflection angle.
Since the regions $u\le 0$ and $\Delta u\le u$ 
are geometrically identical to the Schwarzschild spacetime,
we can introduce the Schwarzschild coordinates
where the metric becomes
\ba
ds^2 = -f(u,r)dt^2+f^{-1}(u,r)dr^2+r^2(d\theta^2+\sin^2\theta d\phi^2).
\ea
The orthonormal basis is naturally introduced by
$\mathbf{e}_0=-\sqrt{f}\,dt$, 
$\mathbf{e}_1=dr/\sqrt{f}$,
$\mathbf{e}_2=rd\theta$, 
$\mathbf{e}_3=r \sin\theta d\phi$,
and then we can define
the angle $\vt$ between the direction of photon propagation 
and the $r-$constant surface as  measured by the static observer,
which 
is expressed as
\ba
\tan\vt=
-\frac{k_a(\mathbf{e}_1)^a}{k_b(\mathbf{e}_3)^b}=
-\frac{1}{r\sqrt{f}}\,\frac{dr}{d\phi},\label{tantheta}
\ea
where $k^a$ is the tangent vector of the null geodesic.
Applying the formula of the right-hand side for arbitrary $u$,
the quantity $\vt$ can be regarded
as the function of $u$, {\it i.e.} $\vartheta(u)$, and
we denote $\vt_0=\vt(0)$ and
introduce the deflection angle by $\Delta\vt = \vt(\Delta u)-\vt_0$.
In this Appendix, we would like to evaluate the value of $\Delta \vt$
caused by the gravitational field
of the null fluid. 
Here, we restrict the value of $\vartheta_0$ to be considered. 
If the photon is emitted exactly in the radial outward direction,
{\it i.e.} $\vartheta_0=-\pi/2$, it certainly arrives at null infinity
because the null fluid never catches up with the photon.
If the photon is emitted exactly in the radial inward direction,
{\it i.e.} $\vartheta_0=\pi/2$, it certainly falls into the black hole.
In what follows, we study what happens for the case $\tan\vartheta_0=\Crz{0}$ as the nontrivial cases.
By this assumption, we exclude the case in which either $|\vartheta_0+\pi/2|$ or $|\vartheta_0-\pi/2|$ is too small.

We begin our analysis of geodesic equations. 
Because of the spherical symmetry of the outgoing Vaidya metric, the angular momentum
\begin{equation}
  L:=r^2\phi'\label{angular_momentum}
\end{equation}
is conserved. The null condition of Eq.~\eqref{Vaidya} gives us
\ba
-fu'^2-2u'r'+\frac{L^2}{r^2}=0.\label{nullcon_Vai}
\ea
The geodesic equation for $r$ \red{and Eqs.~\eqref{angular_momentum} and \eqref{nullcon_Vai} give} us
\ba
r''&=&-\frac{2\dot{M}}{fr}u'r'+\frac{L^2}{r^3}\left(1-\frac{3M}{r}+\frac{\dot{M}}{f}\right).\label{rgeo_Vai}
\ea
In what follows, we express the worldline of a photon by 
$u(\phi)$ and $r(\phi)$. For this reason, we eliminate
the affine parameter $\lambda$ from Eqs.~\eqref{angular_momentum}--\eqref{rgeo_Vai}.
The result is
\begin{equation}
f\red{\left(u_{,\phi}\right)^2}+2u_{,\phi}r_{,\phi} \ = \ r^2,\label{Vaidya-Eq-WL-1}
\end{equation}
\begin{equation}
r_{,\phi\phi} \ =\  \frac{2}{r}\red{\left(r_{,\phi}\right)^2}+\frac{\dot{M}}{fr}\left(r^2-2u_{,\phi}r_{,\phi}\right)+r-3M,\label{Vaidya-Eq-WL-2}
\end{equation}
\red{where we used the the geodesic equation for $\phi$ which is given by
\ba
\phi''&=&-\frac{2}{r}r'\phi'.
\ea
}Since $M(u)$ at the position of the photon can be 
regarded as the function of $\phi$ as $M(u(\phi))$, Eq.~\eqref{Vaidya-Eq-WL-2}
can be rewritten as
\begin{equation}
  r_{,\phi\phi} \ =\  \frac{2}{r}\red{\left(r_{,\phi}\right)^2}+\frac{M_{,\phi}}{r}u_{,\phi}
  +r-3M,\label{Vaidya-Eq-WL-3}
\end{equation}
where we used Eq.~\eqref{Vaidya-Eq-WL-1} and $\dot{M}u_{,\phi}=M_{,\phi}$.
Equations \eqref{Vaidya-Eq-WL-1} and \eqref{Vaidya-Eq-WL-3}
are the basic equations to be studied below.

From Eq.~\eqref{tantheta}, the initial condition
of $r_{,\phi}$ is
\begin{equation}
  \left.r_{,\phi}\right|_{0} = -r_0\tan\vartheta_0 + \Crz{0}.
\end{equation}
\red{We will restrict to the range of $\phi$} where
\begin{equation}
  r_{,\phi} -\left.r_{,\phi}\right|_0 \ =\  \Crz{0}
  \label{Condition-for-the-range}
\end{equation}
holds after the emission\red{, which will be validiated later}. In this range,
\begin{equation}
  r \ = \ r_0-(r_0\phi)\tan\vartheta_0 + \Crz{0}\phi
\end{equation}
holds. 
From Eq.~\eqref{Vaidya-Eq-WL-1},
we have
\begin{eqnarray}
  u_{,\phi} & = & -\frac{r_{,\phi}}{f}+\sqrt{\frac{r^2}{f}+\frac{\red{\left(r_{,\phi}\right)^2}}{f^2}}
  \nonumber\\
  & = & \left.\red{\left[-r_{,\phi}+\sqrt{r^2+{\red{\left(r_{,\phi}\right)^2}}}\right]}\right|_0
  +\Crz{0}
  +\Crz{1}\phi
  \nonumber\\
  & = & \left(\frac{1+\sin\vartheta_0}{\cos\vartheta_0}\right)r_0
  +\Crz{0}
  +\Crz{1}\phi.
  \label{Formula-for-u_phi}
\end{eqnarray}
Then, Eq.~\eqref{Vaidya-Eq-WL-3} implies that
\begin{equation}
r_{,\phi\phi} \ =\  \left.\red{\left[\frac{2}{r}\red{\left(r_{,\phi}\right)^2}
+r\right]}\right|_0 +\Crz{0}+\Crz{1}\phi
+M_{,\phi}\left(\left.\frac{u_{,\phi}}{r}\right|_0+\Crz{-1}+\Crz{0}\phi\right).
\label{formula-for-r_phiphi}
\end{equation}
Integrating this equation, we obtain
\begin{equation}
r_{,\phi} \ =\  \left.r_{,\phi}\right|_0+\left.\red{\left[\frac{2}{r}\red{\left(r_{,\phi}\right)^2}
+r\right]}\right|_0 \phi +\Crz{0}\phi+\Crz{1}\phi^2
+\left( M - M_0 \right) \left(\left.\frac{u_{,\phi}}{r}\right|_0+\Crz{-1}+\Crz{0}\phi\right).
\end{equation}
Comparing this equation with Eq.~\eqref{Condition-for-the-range},
the necessary and sufficient condition for the formula
of Eq.~\eqref{Condition-for-the-range}
to be satisfied is $\phi=\Crz{-1}$.
It would be expected that $\phi$ can be small if we consider short duration. 
Integrating Eq.~\eqref{Formula-for-u_phi},
we have 
\begin{equation}
u \ =\ \left(\frac{1+\sin\vartheta_0}{\cos\vartheta_0}\right)r_0\phi + \Crz{0}\phi,
\end{equation}
and the condition $\phi=\Crz{-1}$ is certainly equivalent to $u=\Crz{0}$.

We now consider the moment $u=\Delta u$ when the
photon goes out from the null fluid region.
\red{Since $\Delta u$ is independent of $r_0$, we have $\Delta u=\Crz{0}$, and} all of the above formulas can be
applied in the range $0\le u\le \Delta u$.
In order to evaluate the deflection angle,
it is better to rewrite Eq.~\eqref{formula-for-r_phiphi}
as
\begin{equation}
  \left(\frac{r_{,\phi}}{r}\right)_{,\phi}
  \ = \
  \left.\red{\left[1+\frac{\red{\left(r_{,\phi}\right)^2}}{r^2}\right]}\right|_0 +\Crz{-1}
+M_{,\phi}\left(\left.\frac{u_{,\phi}}{r^2}\right|_0+\Crz{-2}\right).
\end{equation}
Integrating this equation in the domain $0\le \phi\le \Delta \phi$,
we have
\begin{equation}
  \Delta \left(\frac{r_{,\phi}}{r}\right) \ =\
  \frac{1}{\cos\vartheta_0(1+\sin\vartheta_0)}
  \left(\frac{\Delta u}{r_0}\right) 
  +\frac{1+\sin\vartheta_0}{\cos\vartheta_0}
  \left(\frac{\Delta M}{r_0}\right)+\Crz{-2}.
\end{equation}
Since the angle $\vartheta$ is determined
by Eq.~\eqref{tantheta}, we need to estimate the
quantity $\Delta\left({r_{,\phi}}/{r\sqrt{f}}\right)$.
Here, we have to take account of the change in $f$
as $\Delta f = -2\Delta M/r_0+\Crz{-2}$, and we have
\begin{eqnarray}
  \Delta\left(\frac{r_{,\phi}}{r\sqrt{f}}\right)
  &=& \Delta\left(\frac{r_{,\phi}}{r}\right)
  +\left.\left(\frac{r_{,\phi}}{r}\right)\right|_0\left(\frac{\Delta M}{r_0}\right)
  + \Crz{-2}
  \nonumber\\
  &=&  \frac{1}{\cos\vartheta_0(1+\sin\vartheta_0)}
  \left(\frac{\Delta u}{r_0}\right) 
  +\frac{1}{\cos\vartheta_0}
  \left(\frac{\Delta M}{r_0}\right)+\Crz{-2}.
\end{eqnarray}
Using $\Delta(\tan\vartheta) = \Delta\vartheta/\cos^2\vartheta_0+\Crz{-2}$, 
the deflection angle $\Delta\vartheta$ is
calculated as
\begin{eqnarray}
  \Delta\vartheta 
  & = & -\cos^2\vartheta_0\, \Delta\left(\frac{r_{,\phi}}{r\red{\sqrt{f}}}\right) + \Crz{-2}\nonumber\\
  & = & -\frac{\cos\vartheta_0}{1+\sin\vartheta_0}
  \left(\frac{\Delta u}{r_0}\right) 
  +\cos\vartheta_0
  \left|\frac{\Delta M}{r_0}\right|+\Crz{-2}.
  \label{Deflection-result}
\end{eqnarray}
This is the main result of this Appendix.
The leading order of the deflection angle 
does not depend on the specific form of $M(u)$ and
is just determined by the duration of the null-matter flow, $\Delta u$,
and the change in the mass, $\Delta M$. 
Note that $\Delta\vartheta$ becomes $\Crz{-1}$.

The interpretation of this result is as follows.
The first term of Eq.~\eqref{Deflection-result}
appears without gravity effects and 
reflects the curvature of $r$--constant surfaces, 
or it can be interpreted as the centrifugal force.
If a photon is emitted
in the tangential direction of the $r$--constant surface
in a flat spacetime, it becomes outwardly directed
in propagating along a straight line.
For this reason, this term gives a negative contribution. 
The second term of Eq.~\eqref{Deflection-result}
reflects the gravitational effect of the null matter. 
Since the second term is positive,
the null fluid tends to deflect the
photon in the inward direction. 
Ignoring the term of $\Crz{-2}$, 
the maximum possible deflection angle 
is realized as
\begin{equation}
\Delta\vartheta_{\rm max} \ = \ 
  \frac{M_0}{r_0},
\end{equation}
by adopting $\vartheta_0=0$
and taking the limit $\Delta u\to 0$
and $\Delta M\to M_0$ (assuming the non-negativity of the Bondi mass). \red{Note that in the case of $\Delta u\ll |\Delta M|\approx M_0$, 
the positive definiteness of $\Omega_{IJ}$ of Eq.~\eqref{Def:Omega_IJ} is obviously violated
since $\dot{h}_{IJ}=0$ and $\dot{m}=\lim_{\Delta u\to0}\left(2\Delta M/\Delta u\right)\ll-2$ hold.}
Hence, the deflection angle obtained here \red{in the case with negative definiteness of $\Omega_{IJ}$}
gives an example of the case that a photon emitted
in the tangential direction of an $r$--constant surface \red{will be bent in the inward direction}.
However, it has turned out that the deflection angle is $\Crz{-1}$ and for large $r_0$ it is not
so large  in the current example. Evaluation of the
deflection angle in the case of gravitational waves violating
the positivity of $\Omega_{IJ}$ would be an interesting remaining issue.

Finally, let us discuss whether the photon reaches future null infinity.
For this purpose, 
we consider the impact parameter $b$
just after the photon passes through the null fluid region,
which is given by
$b:=L/E$ with the energy $E:=-k_u=fu^\prime\red{+}r^\prime$.
The value of $b$ is calculated as
\begin{equation}
b \ =\ \frac{r^2}{fu_{,\phi}+r_{,\phi}}
\ =\ r_0\cos\vartheta_0 + \Crz{0}.
\end{equation}
Recalling the assumption $\tan\vartheta_0=\Crz{0}$, 
the value of $b$ is large, $b^{-1}=\Crz{-1}$.
Therefore, the value of $b$ satisfies 
the condition for a photon to reach future null infinity in the Schwarzschild spacetime, $|b|>3\sqrt{3}(M+\Delta M)=\Crz{0}$ (e.g., Sec.~2.9 of \cite{Frolov:1998}).
In order to make the photon fall into the
central black hole, 
we have to choose large $\vartheta_0>0$ such that $\tan\vartheta_0$ does not satisfy $\tan\vartheta_0=\Crz{0}$\red{, 
or to choose small enough $r_0$}.

In conclusion, in the outgoing Vaidya spacetime with an $\Crz{0}$
duration of the null-matter flow, the deflection angle $\Delta\vartheta$
of a photon propagating in the null fluid region can be
explicitly calculated as
Eq.~\eqref{Deflection-result}, and it takes a small value 
of $\Crz{-1}$. 
Therefore,
all photons emitted in the directions which satisfy $\tan\vartheta_0=\Crz{0}$
reach future null infinity without the assumptions
in the Proposition of Sec.~\ref{proof}.


%
%


\begin{thebibliography}{99}

  \bibitem{Amo:2021gcn}
  M.~Amo, K.~Izumi, Y.~Tomikawa, H.~Yoshino and T.~Shiromizu,
  ``Asymptotic behavior of null geodesics near future null infinity: Significance of gravitational waves,''
  Phys. Rev. D \textbf{104}, 064025 (2021).


  \bibitem{Amo:2021rxr}
  M.~Amo, T.~Shiromizu, K.~Izumi, H.~Yoshino and Y.~Tomikawa,
 ``Asymptotic behavior of null geodesics near future null infinity. II. Curvatures, photon surface, and dynamically transversely trapping surface,''
  Phys. Rev. D \textbf{105}, no.6, 064074 (2022).

  \bibitem{Abbott:2016blz}
  B.~P.~Abbott \textit{et al.} (LIGO Scientific and Virgo Collaborations),
  ``Observation of Gravitational Waves from a Binary Black Hole Merger,''
  Phys. Rev. Lett. \textbf{116}, 061102 (2016).
  
  \bibitem{Akiyama:2019cqa}
K.~Akiyama \textit{et al.} (Event Horizon Telescope Collaboration),
``First M87 Event Horizon Telescope results. I. The shadow of the supermassive black hole,''
Astrophys. J. Lett. \textbf{875}, L1 (2019).

\bibitem{Cunha:2017eoe}
P.V.P.~Cunha, C.~A.~R.~Herdeiro and E.~Radu,
``Fundamental photon orbits: black hole shadows and spacetime instabilities,''
Phys. Rev. D \textbf{96} (2017) no.2, 024039.

\bibitem{Cunha:2018acu}
P.~V.~P.~Cunha and C.~A.~R.~Herdeiro,
``Shadows and strong gravitational lensing: a brief review,''
Gen. Rel. Grav. \textbf{50} (2018) no.4, 42.

\bibitem{Claudel:2000} 
C.~M.~Claudel, K.~S.~Virbhadra, and G.~F.~R.~Ellis,
``The Geometry of photon surfaces,''
J.\ Math.\ Phys.\  {\bf 42}, 818 (2001).

\bibitem{Shiromizu:2017ego}
T.~Shiromizu, Y.~Tomikawa, K.~Izumi and H.~Yoshino,
``Area bound for a surface in a strong gravity region,''
PTEP \textbf{2017}, no.3, 033E01 (2017).

\bibitem{Yoshino:2017gqv}
H.~Yoshino, K.~Izumi, T.~Shiromizu and Y.~Tomikawa,
``Extension of photon surfaces and their area: Static and stationary spacetimes,''
PTEP \textbf{2017}, no.6, 063E01 (2017).

\bibitem{Siino:2019vxh}
M.~Siino,
``Causal concept for black hole shadows,''
Class. Quant. Grav. \textbf{38}, no.2, 025005 (2021).

\bibitem{Yoshino:2019dty}
H.~Yoshino, K.~Izumi, T.~Shiromizu and Y.~Tomikawa,
``Transversely trapping surfaces: Dynamical version,''
PTEP \textbf{2020}, no.2, 023E02 (2020).

\bibitem{Cao:2019vlu}
L.~M.~Cao and Y.~Song,
``Quasi-local photon surfaces in general spherically symmetric spacetimes,''
Eur. Phys. J. C \textbf{81}, 714 (2021).

\bibitem{Siino:2021kep}
M.~Siino, ``Black hole shadow and Wandering null geodesics,'' [arXiv:2107.06551 [gr-qc]].

  \bibitem{IDHODH}
  M.~Amo, K.~Izumi, T.~Shiromizu, Y.~Tomikawa and H.~Yoshino, {\it In preparation}.


  \bibitem{Bondi} 
H.~Bondi, M.~G.~J.~van der Burg and A.~W.~K.~Metzner,
``Gravitational waves in general relativity. VII. Waves from axisymmetric isolated systems,''
Proc. Roy. Soc. Lond. A \textbf{269}, 21-52 (1962).

\bibitem{Sachs} 
R.~K.~Sachs,
``Gravitational waves in general relativity. VIII. Waves in asymptotically flat space-times,''
Proc. Roy. Soc. Lond. A \textbf{270}, 103-126 (1962).

\bibitem{Tanabe:2011es}
K.~Tanabe, S.~Kinoshita and T.~Shiromizu,
``Asymptotic flatness at null infinity in arbitrary dimensions,''
Phys. Rev. D \textbf{84}, 044055 (2011).

\bibitem{Hollands:2003xp}
S.~Hollands and A.~Ishibashi,
``Asymptotic flatness at null infinity in higher dimensional gravity,''
arXiv:hep-th/0311178.

\bibitem{Hollands:2003ie}
S.~Hollands and A.~Ishibashi,
``Asymptotic flatness and Bondi energy in higher dimensional gravity,''
J. Math. Phys. \textbf{46}, 022503 (2005).

\bibitem{Ishibashi:2007kb}
A.~Ishibashi,
``Higher Dimensional Bondi Energy with a Globally Specified Background Structure,''
Classical Quantum Gravity \textbf{25}, 165004 (2008).

\bibitem{Cao:2021mwx}
L.~M.~Cao, L.~Y.~Li and L.~B.~Wu,
``Bound on the rate of Bondi mass loss,''
Phys. Rev. D \textbf{104} (2021) no.12, 124017.

\bibitem{dyson}
F. Dyson, ``Gravitational machines,''
in Interstellar Communication, ed. A.G. Cameron, (New York: Benjamin, 1963), chap 12.

\bibitem{thorne}
K. S. Thorne, ``The Theory of Gravitational Radiation: An Introductory Review,''
in Gravitational Radiation, edited by 
N. Deruelle and T. Piran (North-Holland Company, Amsterdam, New York, Oxford, 1983), p. 1.

\bibitem{Frolov:1998}
  V.~P.~Frolov and I.~D.~Novikov,
  {\it Black Hole Physics: Basic Concepts and New Developments},
  (Kluwer Academic Publishers, Dordrecht and Boston, 1998).






  



\end{thebibliography}
\end{document}